%% file: main.tex
\documentclass[sigconf]{acmart}
\AtBeginDocument{%
  }

\copyrightyear{2026}
\acmYear{2026}
\setcopyright{cc}
\setcctype{by}
\acmConference[CIKM '26]{Proceedings of the 35th ACM International Conference on Information and Knowledge Management}{November 07--11, 2026}{Rome, Italy}
\acmBooktitle{Proceedings of the 35th ACM International Conference on Information and Knowledge Management (CIKM '26), November 07--11, 2026, Rome, Italy}
\acmDOI{10.1145/3799682.3839862}
\acmISBN{979-8-4007-2539-5/2026/11}
\usepackage{amsmath}
\usepackage{amsfonts}
\usepackage{enumitem}
\usepackage{graphicx}
\usepackage{stfloats}
\allowdisplaybreaks

\newcommand{\name}{\textit{CMRec}}
\begin{document}

% \title{Collaboration and Confrontation: Infusing Personalized Knowledge from Large Language Models into Recommendation}
% \title{Masked Diffusion Generative Recommendation}
\title{Cross-Country Code-Mixing for Generative Recommendation}

%% IMPORTANT: author names, order, e-mails, ORCIDs and affiliation (incl. city/country)
%% below must match the ACM rightsreview form EXACTLY, character for character.
%% Authorship order follows the accepted submission and must not be changed.
\author{Yuan Gao}
\authornote{Both authors contributed equally to this research.}
\orcid{0009-0004-2055-9941}
\email{jianbei.gy@alibaba-inc.com}
\affiliation{%
  \institution{Alibaba International Digital Commerce Group}
  \city{Beijing}
  \country{China}}

\author{Hao Deng}
\authornotemark[1]
\orcid{0009-0002-6335-7405}
\email{denghao.deng@alibaba-inc.com}
\affiliation{%
  \institution{Alibaba International Digital Commerce Group}
  \city{Beijing}
  \country{China}}

\author{Haibo Xing}
\orcid{0009-0006-5786-7627}
\email{xinghaibo.xhb@alibaba-inc.com}
\affiliation{%
  \institution{Alibaba International Digital Commerce Group}
  \city{Beijing}
  \country{China}}

\author{Yi Xu}
\orcid{0009-0007-3571-8791}
\email{xy397404@alibaba-inc.com}
\affiliation{%
  \institution{Alibaba International Digital Commerce Group}
  \city{Beijing}
  \country{China}}

\author{Lingyu Mu}
\orcid{0009-0005-4959-0851}
\email{moulingyu.mly@alibaba-inc.com}
\affiliation{%
  \institution{Alibaba International Digital Commerce Group}
  \city{Beijing}
  \country{China}}

\author{Jinxin Hu}
\authornote{Corresponding author.}
\orcid{0000-0002-7252-5207}
\email{jinxin.hjx@alibaba-inc.com}
\affiliation{%
  \institution{Alibaba International Digital Commerce Group}
  \city{Beijing}
  \country{China}}

\author{Yu Zhang}
\orcid{0000-0002-6057-7886}
\email{daoji@alibaba-inc.com}
\affiliation{%
  \institution{Alibaba International Digital Commerce Group}
  \city{Beijing}
  \country{China}}

\author{Xiaoyi Zeng}
\orcid{0000-0002-3742-4910}
\email{yuanhan@taobao.com}
\affiliation{%
  \institution{Alibaba International Digital Commerce Group}
  \city{Hangzhou}
  \country{China}}

\renewcommand{\shortauthors}{Yuan Gao et al.}

\begin{abstract}
Cross-country recommendation on modern e-commerce platforms is typically deployed with disjoint user and item ID spaces across markets, removing the shared anchors that conventional cross-domain methods rely on. Generative recommendation (GR) mitigates this by mapping items into a shared token space and training a unified model, but existing approaches keep behavior sequences strictly country-specific, so knowledge transfer occurs only at the parameter level and remains absent at the data level. Inspired by code-switching corpora in multilingual natural language
processing, we propose \textbf{\name}, a cross-country GR framework that injects cross-country supervision at the data level via dual-constrained, context-aware code-mixing. \name\  first learns a shared semantic codebook from multi-modal content and behavioral co-occurrence across countries. It then uses this codebook to synthesize mixed-country sequences via token-level substitutions that satisfy both static (content) and dynamic (\textit{e.g.}, price, audience, popularity) constraints. Finally, it introduces a context-aware loss that reweights mixed samples according to their plausibility in the current sequence. Experiments on two real-world multi-country datasets and an online A/B test show that \name\  substantially improves recommendation quality in data-sparse countries while preserving performance in data-rich countries, achieving \textbf{+1.77\%} advertising revenue and \textbf{+2.64\%} orders on a large-scale e-commerce platform.
\end{abstract}

%% Generated by the tool at https://dl.acm.org/ccs



\maketitle

\input{0_introduction}

\input{1_preliminary}
\input{2_method}

\input{3_result_and_analysis}

\section{GenAI Usage Disclosure}
Generative AI was used in the preparation of this paper to assist with language polishing, grammar checking, and improving the clarity of the presentation. All technical ideas, including the design of the method, the implementation of experiments, and the interpretation of results, are solely contributed by the authors.
\bibliographystyle{ACM-Reference-Format}
\bibliography{reference}

\end{document}

%% file: 0_introduction.tex
% !TEX root = main.tex
\section{Introduction}
\label{sec:intro}

\begin{figure}[t]
\centering
\includegraphics[width=0.8\columnwidth]{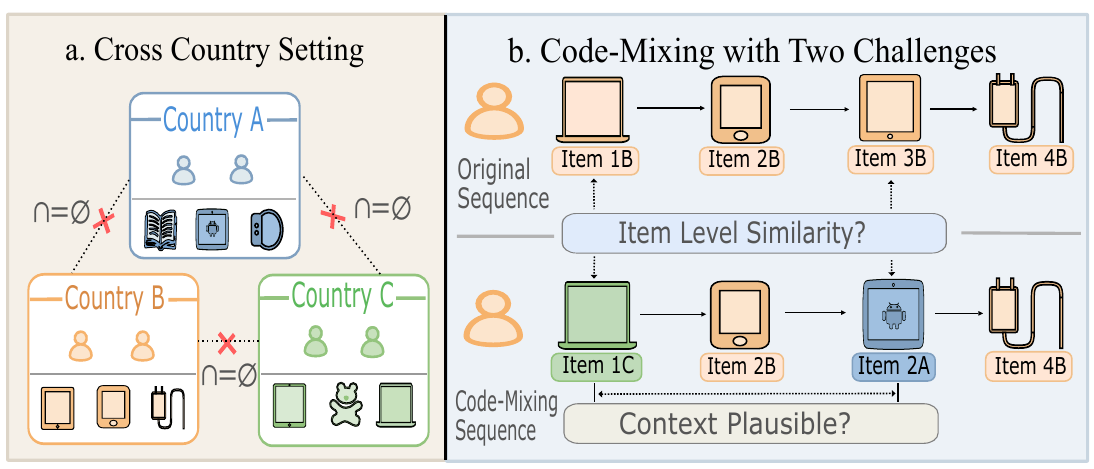}
% \caption{Concept illustration of cross-country code-switching in generative recommendation. Items from different markets are mapped to a shared token space, enabling token-level substitution to synthesize mixed sequences.}
\caption{(a) Cross-country setting. (b) Example and challenges of cross-country behavior-sequence code-mixing.}
\Description{Two-panel schematic. Panel (a) shows the cross-country setting: several country-specific markets with disjoint user and item identifier spaces, whose items are quantized into one shared semantic token space. Panel (b) shows a user behaviour sequence in which one item is replaced by its cross-country counterpart through token-level substitution, together with the two failure cases this can cause: substitutions that violate static content attributes, and substitutions that violate time-varying attributes such as price, audience and popularity.}
\label{fig:concept}
\end{figure}

% === [P1 v1, archived] original first paragraph: setting + imbalance + CDR-failure mixed in one shot ===
% Modern e-commerce platforms increasingly operate across many countries. In this cross-country setting, user and item ID spaces are typically disjoint, and behavioral data are highly imbalanced: a few large countries accumulate rich interaction histories, while smaller ones remain data-sparse. The absence of overlapping users and items removes the bridges that conventional cross-domain recommendation methods rely on and makes it difficult to transfer knowledge explicitly from mature to emerging markets.

% === [P1 v2] soft setting only: universality + light importance hook, no pain-point, no CDR contrast ===
% Modern e-commerce platforms increasingly operate across many countries~\cite{xing2025reg4rec,yang2024not,li2025adaptive,cross-country-survey}. Each country is deployed and logged as an independent system, so user interactions are observed only within a single market. As a result, both user/item ID spaces and behavioral sequences are partitioned by country, with no natural overlap across markets.
Modern e-commerce platforms increasingly operate across many countries~\cite{xing2025reg4rec,yang2024not,li2025adaptive}. In practice, each country is deployed and logged as an independent system: user interactions are recorded only within a single country, and both user and item ID spaces are maintained separately~\cite{cross-country-survey}. Consequently, both IDs and behavior sequences are strictly partitioned by country, with no natural overlap across countries, as illustrated in Figure~\ref{fig:concept} (a).

This setup differs from the well-studied cross-domain recommendation (CDR) setting~\cite{li2025cd,aaai24,zang2022survey,zhang2023collaborative,zhang2025comprehensive}, where knowledge transfer typically relies on some form of overlap across domains, such as shared users, shared items, or manually anchored entities that carry collaborative signals. Recent CDR methods based on generative recommendation (GR), such as GenCDR~\cite{hu2026ids} and GMC~\cite{jin2025generative}, relax this ID-level requirement by mapping items from different domains into a shared semantic token space and training a unified generative model. In this way, knowledge transfer occurs at the \textbf{parameter level}, mediated by the shared backbone and token space. However, GR learns primarily from item co-occurrence within training sequences~\cite{hou2025generative,gr_survey,RCLRec}. In the cross-country setting, these sequences remain strictly partitioned by country, so items from one country never appear in another country's context, leaving cross-country reinforcement absent at the \textbf{data level}. Consequently, rich interaction signals in large markets cannot benefit small ones, and the potential for cross-country knowledge sharing remains largely untapped.

A natural reference comes from multilingual natural language processing (NLP), where models learn cross-lingual alignment from code-switched corpora, in which tokens from different languages appear in the same training sequence~\cite{blevins2022contamination,briakou2023incidental,li2024prealign}. Motivated by this, we investigate whether an analogous construction is possible for cross-country GR, by building mixed-country sequences that interleave items from different markets, as illustrated in Figure~\ref{fig:concept} (b). However, no such corpus exists naturally; it must be \textbf{synthesized by substitution}. This raises a recommendation-specific question: \emph{what counts as a valid substitute?} Existing substitution-based augmentation methods~\cite{sracl,missrec,LLM4CDSR}, mostly developed for CDR settings, match items by content (\textit{e.g.}, titles, descriptions, images) and replace them in user sequences. Such recipes can in principle be applied to cross-country GR, but they fall short along two complementary axes. \textbf{(a) Substitution criteria are incomplete at the item level.} Content similarity alone ignores two critical signals: \emph{behavioral} (\textit{e.g.}, co-purchase and co-click patterns) and \emph{market} (\textit{e.g.}, price, audience, popularity). Two items that look similar can play very different roles across markets~\cite{cross-country-survey}, so a content-only criterion injects substantial noise. \textbf{(b) Substitution granularity is too coarse at the context level.} Even if a candidate is reasonable in isolation, it may still conflict with the user's current intent. For example, in a sequence dominated by Apple devices, swapping an item with an Android pad that looks similar in content and basic attributes is syntactically plausible but contextually wrong, and this cannot be detected by any offline item-pair filter.
To address these limitations, we propose \textbf{\name}, a cross-country GR framework that injects cross-country supervision at the data level via dual-constrained, context-aware code-mixing. \name\  first learns a shared semantic codebook from multi-modal content and behavioral co-occurrence across countries, which serves as both the GR tokenizer and a common semantic space for cross-country substitution. On top of this codebook, we perform token-level, dual-constrained code-mixing: substitutions are applied to semantic token sequences rather than country-specific item IDs, and candidates must satisfy both static (content) and dynamic (\textit{e.g.}, price, audience, popularity) constraints, enabling finer-grained, market-aware cross-country mixing. Finally, we introduce a context-aware loss that lets the GR model reweight mixed samples according to their plausibility in the current context, downweighting noisy substitutions while preserving useful cross-country signals. We conduct experiments on two real-world multi-country datasets and an online A/B test, achieving improvements of \textbf{+1.77\% advertising revenue} and \textbf{+2.64\% orders}, demonstrating the practical effectiveness of \name\  in production.

%% file: 1_preliminary.tex
% !TEX root = main.tex
\section{Problem Formulation}
We consider cross-country recommendation over a country set $\mathcal{C}$, where each $c \in \mathcal{C}$ is an independent domain with disjoint user and item sets $\mathcal{U}^{(c)}$ and $\mathcal{I}^{(c)}$ (i.e., $\mathcal{U}^{(c)} \cap \mathcal{U}^{(c')} = \mathcal{I}^{(c)} \cap \mathcal{I}^{(c')} = \emptyset$ for $c \neq c'$). For each user $u \in \mathcal{U}^{(c)}$, we denote the interaction sequence as $x_c = (i_1, \dots, i_T)$ with $i_t \in \mathcal{I}^{(c)}$, and the target item as $y_c \in \mathcal{I}^{(c)}$. Following the GR paradigm~\cite{tiger,xing2025reg4rec}, we map each item to a discrete token sequence via a codebook and formulate recommendation as conditional sequence generation in this unified token space. Let $\mathcal{D}^{(c)} = \{(x_c, y_c)\}$ collect the per-country training pairs from users $u \in \mathcal{U}^{(c)}$. For joint multi-country training, we pool samples from all countries into a unified set $\mathcal{D} = \bigcup_{c \in \mathcal{C}} \mathcal{D}^{(c)} = \{(x, y)\}$, on which the GR model is trained to maximize
\begin{equation}
\mathcal{L}_{\text{GR}}(\theta) 
= \sum_{(x, y) \in \mathcal{D}} \log P_\theta(y \mid x),
\end{equation}
where $\theta$ are the model parameters.

%% file: 2_method.tex
\section{Method}

\begin{figure*}[!t]
\centering
\includegraphics[width=0.8\textwidth, trim=0 10pt 0 10pt, clip]{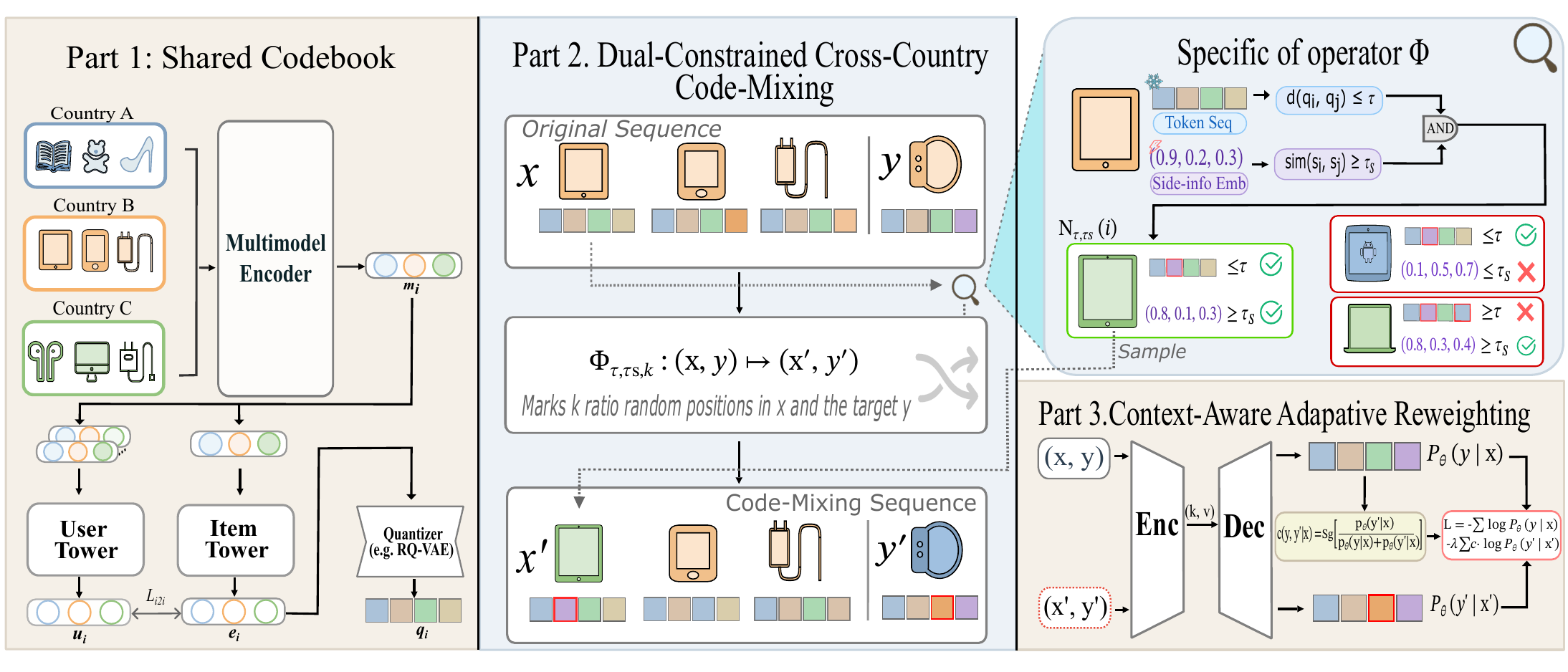}
\caption{Overview of \name\ . \textbf{Part 1} learns a shared semantic codebook from multimodal content and behavioral i2i co-occurrence, producing token sequences $\mathbf{q}_i$ that are used by the GR model. \textbf{Part 2} uses the codebook plus side-info to select cross-country neighbors and grafts them into the user sequence, yielding a code-mixed twin $(\mathbf{x}',y')$. \textbf{Part 3} lets the GR model itself reweight each twin by its plausibility under the original context, downweighting noisy substitutions during training.}
\label{fig:method}
\Description{Three-part pipeline diagram of the proposed framework. Part 1 trains a shared semantic codebook from multimodal item content and behavioural item-to-item co-occurrence, turning every item of every country into a token sequence that the generative recommendation model consumes. Part 2 uses that codebook together with side information to pick cross-country neighbour items and graft them into a user behaviour sequence, producing a code-mixed twin sequence and its rewritten target. Part 3 feeds both the original and the mixed sequence to the generative recommendation model itself, which scores their plausibility under the same user context and reweights each mixed sample so that noisy substitutions contribute less to the training loss.}
\end{figure*}

We train a unified cross-country GR model that shares information across countries while respecting country-specific dynamics. To this end, \name\  comprises three components: (i) a shared semantic codebook that tokenizes items from all countries using behavior and content signals; (ii) a dual-constrained code-mixing mechanism that substitutes tokens across countries under static and dynamic constraints; and (iii) a context-aware loss reweighting scheme that downweights noisy mixed samples during training.
% We aim to train a unified cross-country GR model that can share information across markets while respecting country-specific dynamics. To this end, TNSRec builds on three components: (i) a shared semantic codebook that tokenizes items from all countries in a behavior-aware and content-aware way; (ii) a dual-constrained code-mixing mechanism that performs token-level cross-country substitution under static and dynamic constraints; and (iii) a context-aware twin reweighting scheme that lets the GR model downweight noisy mixed samples during training.

\subsection{Behavior- and Content-Grounded Shared Codebook}
\label{sec:codebook}

To enable a unified cross-country GR backbone, we first construct a shared semantic codebook that is informed by both multi-modal content and behavioral signals. This codebook serves two roles at once: it tokenizes items for GR and provides a common semantic space for cross-country token substitution.

We first encode each item's multimodal side information (\textit{e.g.}, title, image) into a content embedding $\mathbf{m}_i$ using a pretrained model~\cite{bai2025qwen25vl} shared across countries. This provides a unified semantic space where items from all countries are comparable at the content level. However, content alone does not capture how items actually behave in each country: items with similar visuals can have very different popularity or co-occurrence patterns, and these behavior signals are implicitly embedded in user interaction sequences.~\cite{mcauley2015styles,yuan2023idvsmodality}

To inject such behavioral information, we further train a single behavior-driven item-to-item (i2i) alignment model jointly across countries. On top of $\mathbf{m}_i$, we adopt a two-tower architecture with sampled softmax~\cite{deng2025csmf,xu2022mixture}: the user tower maps a behavior sequence to a user embedding $\mathbf{u}_i$, and the item tower maps $\mathbf{m}_i$ to an item embedding $\mathbf{e}_i = T(\mathbf{m}_i)$. The joint i2i training objective is 
\begin{equation}
\mathcal{L}_{\text{i2i}} = -\sum_{i} \log \frac{\exp(\mathbf{u}_i^{\top}\mathbf{e}_{i^{+}})}{\exp(\mathbf{u}_i^{\top}\mathbf{e}_{i^{+}}) + \sum_{j \in \mathcal{N}_i} \exp(\mathbf{u}_i^{\top}\mathbf{e}_j)},
\end{equation}
where the negative pool $\mathcal{N}_i$ is shared across countries~\cite{huang2020embedding}, so $\mathbf{e}_i$ aligns items that are both content-similar and behaviorally similar in user interaction patterns.

We then apply a residual quantizer (\textit{e.g.}, RQ-VAE~\cite{tiger}) to transform $\{\mathbf{e}_i\}$ into shared token sequences $\mathbf{q}_i = (q_i^{(1)},\dots,q_i^{(M)})$. These tokens form a unified codebook that we use both as the GR tokenizer and as the substrate for cross-country code-mixing.

\subsection{Dual-Constrained Cross-Country Code-Mixing}
\label{sec:mixing}
With the shared codebook in place, items from different countries can be matched in a common token space. We now introduce cross-country code-mixing at the token level to expose GR to mixed sequences across countries.

A naive approach would substitute items based only on their codebook representations. However, the codebook is updated only infrequently and does not capture fast-changing attributes such as price discount, popularity, and audience tags, which are crucial for deciding whether two items are interchangeable in practice. To address this, we adopt a dual-constrained filter. Each item $i$ is equipped with a side-information embedding $\mathbf{s}_i$ over these dynamic attributes, with each element discretized by per-country quantile bucketization so that values from different countries are placed on a comparable scale, and we define the cross-country candidate set as
\begin{equation}
\hat{\mathcal{N}}_{\tau,\tau_s}(i) = \Big\{ j \in \!\!\bigcup_{c' \neq c}\!\! \mathcal{I}^{(c')} \,\Big|\, d(\mathbf{q}_{i},\mathbf{q}_j) \le \tau \;\land\; \mathrm{sim}(\mathbf{s}_i,\mathbf{s}_j) \ge \tau_s \Big\},
\end{equation}
where $d$ is the Hamming distance~\cite{chan2020approximating} between token sequences and $\mathrm{sim}$ is cosine similarity, enforcing static and dynamic consistency.

Given a training pair $(\mathbf{x},y)$, the operator $\Phi_{\tau,\tau_s,k}: (\mathbf{x},y) \mapsto (\mathbf{x}',y')$ replaces the items at a rate-$k$ fraction of slots in $\mathbf{x}$ and $y$ with neighbors drawn from $\hat{\mathcal{N}}_{\tau,\tau_s}(\cdot)$, leaving the rest intact while injecting cross-country co-occurrence at the token level. We then apply $\Phi$ to a rate-$p$ subsample of $\mathcal{D}$ to form the twin set $\mathcal{D}'$, with $p$ controlling the cross-country mixing strength.

\subsection{Context-Aware Adaptive Reweighting}
\label{sec:reweight}

The filter $\hat{\mathcal{N}}_{\tau,\tau_s}$ ensures item-level plausibility, but it is blind to the user context and to how the model’s predictions evolve during training. Some synthetic substitutions can still be incompatible with the surrounding sequence, especially in long and noisy histories. We therefore let the GR model itself assess each substituted item during training.
% The offline filter $\hat{\mathcal{N}}_{\tau,\tau_s}$ ensures catalog-level plausibility, but it is blind to the user context and to the evolving beliefs of the GR model. Some synthetic twins can still be incompatible with the surrounding sequence, especially in long and noisy histories. We therefore let the GR model itself assess each twin during training.
For a substituted example $(\mathbf{x}', y')$ derived from an original pair $(\mathbf{x}, y)$, we define the context-conditional substitutability as
% For a twin $(\mathbf{x}', y')$ derived from an original pair $(\mathbf{x}, y)$, we define the context-conditional substitutability as
\begin{equation}
w(y, y' \mid \mathbf{x}) = \mathrm{sg}\!\left[\frac{P_\theta(y' \mid \mathbf{x})}{P_\theta(y \mid \mathbf{x}) + P_\theta(y' \mid \mathbf{x})}\right] \in [0,1],
\end{equation}
% where $P_\theta(\cdot \mid \mathbf{x})$ is the autoregressive item likelihood under the original context and $\mathrm{sg}[\cdot]$ denotes stop-gradient, so $c$ informs the loss but does not receive gradients. We deliberately evaluate $y'$ under the original context $\mathbf{x}$ rather than the mixed $\mathbf{x}'$: $P_\theta$ is better calibrated on the observed $\mathbf{x}$, which keeps $c$ from being driven by the model's unstable response to out-of-distribution synthetic inputs while its beliefs are still evolving.
where $P_\theta(\cdot \mid \mathbf{x})$ is the autoregressive item likelihood under the original context and $\mathrm{sg}[\cdot]$ denotes stop-gradient, so $w$ affects the loss but does not receive gradients. We deliberately evaluate $y'$ under the original context $\mathbf{x}$ rather than the mixed $\mathbf{x}'$, since $P_\theta$ is better calibrated on the observed $\mathbf{x}$; this prevents $w$ from being dominated by the model’s unstable response to out-of-distribution synthetic substitutions early in training.
% where $P_\theta(\cdot \mid \mathbf{x})$ is the token likelihood under the original context and $\mathrm{sg}[\cdot]$ denotes stop-gradient, so $c$ affects the loss but does not receive gradients. We deliberately evaluate $y'$ under the original context $\mathbf{x}$ rather than the mixed $\mathbf{x}'$, since $P_\theta$ is better calibrated on the observed $\mathbf{x}$; this prevents $c$ from being dominated by the model’s unstable response to out-of-distribution synthetic substitutions early in training. This design deliberately decouples the assessment of substitution quality (via $c$ under $\mathbf{x}$) from the optimization of the GR model (via $\mathcal{L}$ under $\mathbf{x}'$), preventing feedback loops and ensuring stable training dynamics.

We then combine original and substituted supervision in the training objective:
\begin{equation}
\mathcal{L}(\theta) = -\sum_{(\mathbf{x},y) \in \mathcal{D}} \log P_\theta(y \mid \mathbf{x}) - \lambda \sum_{(\mathbf{x}',y') \in \mathcal{D}'} w(y, y' \mid \mathbf{x}) \log P_\theta(y' \mid \mathbf{x}'),
\end{equation}
where $\lambda$ controls the overall contribution of substituted examples. Because $w$ is derived from the model's own predictions, it acts as a context-aware weight: early in training most substitutions receive similar weights, while later the model downweights contextually invalid substitutes that the item-level filter cannot distinguish.

%% file: 3_result_and_analysis.tex
% z% !TEX root = main.tex
\section{Experiments}
\begin{table*}[t]
\centering
\caption{Performance comparison on the industrial and Amazon M2 datasets. Each metric reports the 6-country/locale average (Avg6) plus three representative ones, with superscripts $L$/$M$/$S$ marking large/medium/small volume (Amazon M2 has no $M$ tier). Best results are in \textbf{bold} and second-best are \underline{underlined}. ``\textbf{Improv.}'' shows the relative improvement (\%) over the best baseline.}
\label{tab:deep_metrics}
\scriptsize
\setlength{\tabcolsep}{2pt}
\begin{tabular*}{0.95\textwidth}{@{\extracolsep{\fill}} l cccc |cccc |cccc |cccc @{}}
\toprule
\textbf{Method}
& \multicolumn{4}{c}{\textbf{Recall@10}}
& \multicolumn{4}{c}{\textbf{Recall@100}}
& \multicolumn{4}{c}{\textbf{NDCG@10}}
& \multicolumn{4}{c}{\textbf{NDCG@100}} \\
% \cmidrule(lr){2-5} \cmidrule(lr){6-9} \cmidrule(lr){10-13} \cmidrule(lr){14-17} % 删除这一行
\midrule
\midrule
\multicolumn{17}{c}{\emph{(a) Industrial Dataset (6 countries)}}\\
\midrule
 & Avg6 & A$^{L}$ & B$^{M}$ & C$^{S}$
 & Avg6 & A$^{L}$ & B$^{M}$ & C$^{S}$
 & Avg6 & A$^{L}$ & B$^{M}$ & C$^{S}$
 & Avg6 & A$^{L}$ & B$^{M}$ & C$^{S}$ \\
\cmidrule(lr){2-17}
SASRec         & 0.1882 & 0.1993 & 0.1867 & 0.1705 & 0.2648 & 0.2807 & 0.2632 & 0.2404 & 0.0941 & 0.0995 & 0.0933 & 0.0853 & 0.1458 & 0.1546 & 0.1448 & 0.1322 \\
HSTU           & 0.2117 & 0.2244 & 0.2098 & 0.1929 & 0.2983 & 0.3158 & 0.2957 & 0.2725 & 0.1058 & 0.1122 & 0.1049 & 0.0964 & 0.1642 & 0.1738 & 0.1629 & 0.1498 \\
TIGER          & 0.2261 & 0.2396 & 0.2243 & 0.2062 & 0.3187 & 0.3377 & 0.3163 & 0.2909 & 0.1131 & 0.1198 & 0.1122 & 0.1031 & 0.1753 & 0.1858 & 0.1739 & 0.1600 \\
REG4Rec        & 0.2372 & 0.2517 & 0.2354 & 0.2163 & 0.3342 & 0.3537 & 0.3318 & 0.3057 & 0.1187 & 0.1258 & 0.1177 & 0.1082 & 0.1838 & 0.1949 & 0.1824 & 0.1681 \\
GenCDR         & \underline{0.2423} & 0.2493 & \underline{0.2407} & 0.2247 & 0.3413 & 0.3562 & \underline{0.3393} & 0.3168 & \underline{0.1212} & 0.1246 & \underline{0.1204} & 0.1123 & 0.1877 & 0.1961 & \underline{0.1866} & 0.1743 \\
GenCDR+Cont    & 0.2404 & 0.2456 & 0.2371 & \underline{0.2311} & \underline{0.3432} & 0.3496 & 0.3359 & \underline{0.3248} & 0.1202 & 0.1227 & 0.1186 & \underline{0.1156} & \underline{0.1891} & 0.1928 & 0.1847 & \underline{0.1793} \\
GenCDR+CF      & 0.2384 & \underline{0.2528} & 0.2378 & 0.2256 & 0.3412 & \underline{0.3575} & 0.3387 & 0.3179 & 0.1182 & \underline{0.1264} & 0.1192 & 0.1109 & 0.1871 & \underline{0.1969} & 0.1853 & 0.1768 \\
\textbf{\name} & \textbf{0.2496} & \textbf{0.2567} & \textbf{0.2494} & \textbf{0.2434} & \textbf{0.3613} & \textbf{0.3672} & \textbf{0.3567} & \textbf{0.3474} & \textbf{0.1236} & \textbf{0.1278} & \textbf{0.1239} & \textbf{0.1198} & \textbf{0.1972} & \textbf{0.2018} & \textbf{0.1957} & \textbf{0.1919} \\
% \cmidrule(lr){1-17}
\midrule
\textbf{Improv.}    & +3.01\% & +1.54\% & +3.61\% & +5.32\% & +5.27\% & +2.71\% & +5.13\% & +6.96\% & +1.98\% & +1.11\% & +2.91\% & +3.63\% & +4.28\% & +2.49\% & +4.88\% & +7.03\% \\
\midrule
\midrule
\multicolumn{17}{c}{\emph{(b) Amazon M2 Dataset (6 locales)}}\\
\midrule
 & Avg6 & UK$^{L}$ & FR$^{S}$ & ES$^{S}$
 & Avg6 & UK$^{L}$ & FR$^{S}$ & ES$^{S}$
 & Avg6 & UK$^{L}$ & FR$^{S}$ & ES$^{S}$
 & Avg6 & UK$^{L}$ & FR$^{S}$ & ES$^{S}$ \\
\cmidrule(lr){2-17}
SASRec         & 0.2774 & 0.2293 & 0.3028 & 0.3095 & 0.4823 & 0.3986 & 0.5274 & 0.5389 & 0.1388 & 0.1147 & 0.1516 & 0.1548 & 0.2653 & 0.2192 & 0.2898 & 0.2963 \\
HSTU           & 0.3212 & 0.2654 & 0.3506 & 0.3584 & 0.5584 & 0.4618 & 0.6099 & 0.6236 & 0.1606 & 0.1327 & 0.1753 & 0.1792 & 0.3070 & 0.2539 & 0.3354 & 0.3428 \\
TIGER          & 0.3406 & 0.2816 & 0.3719 & 0.3803 & 0.5926 & 0.4898 & 0.6471 & 0.6616 & 0.1703 & 0.1408 & 0.1861 & 0.1902 & 0.3258 & 0.2694 & 0.3559 & 0.3638 \\
REG4Rec        & 0.3523 & 0.2916 & 0.3846 & 0.3931 & 0.6124 & 0.5066 & 0.6688 & 0.6843 & 0.1762 & 0.1458 & 0.1921 & 0.1964 & 0.3369 & 0.2786 & 0.3679 & 0.3764 \\
GenCDR         & \underline{0.3603} & 0.2894 & \underline{0.3928} & \underline{0.4009} & 0.6266 & 0.5186 & 0.6839 & 0.6994 & \underline{0.1798} & 0.1447 & \underline{0.1963} & \underline{0.2004} & 0.3446 & 0.2852 & 0.3761 & 0.3848 \\
GenCDR+Cont    & 0.3579 & 0.2862 & 0.3907 & 0.3991 & \underline{0.6291} & 0.5149 & \underline{0.6893} & \underline{0.7063} & 0.1784 & 0.1432 & 0.1951 & 0.1997 & \underline{0.3463} & 0.2831 & \underline{0.3793} & \underline{0.3884} \\
GenCDR+CF      & 0.3559 & \underline{0.2922} & 0.3884 & 0.3961 & 0.6271 & \underline{0.5198} & 0.6862 & 0.7035 & 0.1764 & \underline{0.1465} & 0.1942 & 0.1985 & 0.3443 & \underline{0.2860} & 0.3781 & 0.3865 \\
\textbf{\name} & \textbf{0.3748} & \textbf{0.2997} & \textbf{0.4137} & \textbf{0.4203} & \textbf{0.6671} & \textbf{0.5379} & \textbf{0.7432} & \textbf{0.7588} & \textbf{0.1847} & \textbf{0.1484} & \textbf{0.2033} & \textbf{0.2068} & \textbf{0.3637} & \textbf{0.2953} & \textbf{0.4058} & \textbf{0.4136} \\
% \cmidrule(lr){1-17}
\midrule
\textbf{Improv.}    & +4.02\% & +2.57\% & +5.32\% & +4.84\% & +6.04\% & +3.48\% & +7.82\% & +7.43\% & +2.73\% & +1.30\% & +3.57\% & +3.19\% & +5.03\% & +3.25\% & +6.99\% & +6.49\% \\
\bottomrule
\end{tabular*}
\end{table*}

\subsection{Experimental Settings}
\textbf{Datasets \& Evaluation Metrics.} We evaluate \name\ on \textbf{two multi-country datasets} with highly imbalanced data distributions. The first is an in-house industrial advertising dataset from an e-commerce platform. It contains over 1 billion interactions from 19 million users across 25 million advertisements across 6 countries (A--F). The largest country contributes about 10$\times$ more training interactions than the smallest, and user and item IDs are completely disjoint across countries. The second is Amazon M2~\cite{amazonm2}, the public KDD Cup 2023 multilingual session benchmark covering 6 locales (high-resource: UK/DE/JP; low-resource: IT/FR/ES). 
% We mix its task~1 and task~2 data to simulate a cross-country setting. 
The high-resource locales account for roughly 91\% of all training sessions, yielding a pronounced long-tail distribution. For offline evaluation, we report Recall@10/100 and NDCG@10/100~\cite{deng2025csmf,tiger} for each representative country/locale, and also report averages over all countries/locales.

\textbf{Baselines \& implementation.}
% We compare with \textbf{five representative methods}: SASRec~\cite{sasrec}, GenCDR~\cite{hu2026ids}, REG4Rec~\cite{xing2025reg4rec}, HSTU~\cite{zhai2024actions}, and TIGER~\cite{tiger}. To examine the effect of single-signal data augmentation, we further follow SRA-CL~\cite{sracl} and build a content-augmented variant on top of GenCDR (denoted GenCDR+Cont), and follow CoSeRec~\cite{coserec} and TiCoSeRec~\cite{dang2023uniform} to build a collaboration-augmented counterpart grounded in user-behavior signals (denoted GenCDR+CF). Both augmented variants use the same substitution ratio and sampling budget as \name\ for a controlled comparison. 
% All baselines are implemented according to their original designs and trained on the same dataset partition for a fair comparison. 
% For \name, each item's content is first encoded by Qwen-VL-32B~\cite{bai2023qwen} into a 64-d vector and then tokenized by an RQ-VAE~\cite{tiger} into $L=8$ sub-IDs. We adopt GenCDR as the backbone (Figure~\ref{fig:sensitivity} further verifies that our augmentation strategy generalizes to different backbones). For augmentation, we set $p=0.1$ and $k=0.1$, balancing downstream performance and search-time cost. We use $\lambda=0.5$ and adopt relatively strict filtering thresholds $\tau=3$ and $\tau_s=0.8$.
We compare with five representative methods plus two augmented variants: SASRec~\cite{sasrec}, GenCDR~\cite{hu2026ids}, REG4Rec~\cite{xing2025reg4rec}, HSTU~\cite{zhai2024actions}, and TIGER~\cite{tiger}. 
To study single-signal data augmentation, we follow SRA-CL~\cite{sracl} and build a content-augmented variant on top of GenCDR, denoted GenCDR+Cont. We also follow CoSeRec~\cite{coserec} and TiCoSeRec~\cite{dang2023uniform} to build a collaboration-augmented variant based on user-behavior signals, denoted GenCDR+CF. Both augmented variants use the same substitution ratio and sampling budget as \name\ for a controlled comparison. 
All baselines follow their original designs and are trained on the same dataset partition. 
For \name, we first encode each item's content into a 64-dimensional vector using Qwen-VL-32B~\cite{bai2025qwen25vl}. We then tokenize this vector with an RQ-VAE~\cite{tiger} into a token sequence of length $L=4$. We adopt GenCDR as the backbone. We set $p=0.1$ and $k=0.1$ to balance performance and computational cost. We further set $\lambda=0.5$ and use relatively strict filtering thresholds, $\tau=1$ and $\tau_s=0.8$.
% We adopt GenCDR as the backbone 
% Figure~\ref{fig:sensitivity} shows that our augmentation strategy also generalizes to other backbones. 
% For augmentation, 
% and set $p=0.1$ and $k=0.1$ to balance downstream performance and computational cost. We use $\lambda=0.5$ and choose relatively strict filtering thresholds, with $\tau=3$ and $\tau_s=0.8$.

\subsection{Experimental Results}

\begin{table}[!b]
\centering
\caption{Ablation on the Industrial dataset (6-country avg.). Numbers in parentheses are relative gaps vs.\ Full.}
\label{tab:ablation}
\scriptsize
\begin{tabular*}{\columnwidth}{@{\extracolsep{\fill}} lcc @{}}
\toprule
\textbf{Variant} & \textbf{Recall@100} & \textbf{NDCG@100} \\
\midrule
\textbf{\name\ (Full)}            & \textbf{0.3613} & \textbf{0.1972} \\
\,w/o Token-level distance constraint   & 0.3447\,{\scriptsize($-$4.59\%)} & 0.1894\,{\scriptsize($-$3.96\%)} \\
\,w/o Side-info Similarity constraint    & 0.3573\,{\scriptsize($-$1.11\%)} & 0.1947\,{\scriptsize($-$1.27\%)} \\
\,w/o Context-aware Adaptive Reweighting                   & 0.3531\,{\scriptsize($-$2.27\%)} & 0.1928\,{\scriptsize($-$2.23\%)} \\
\bottomrule
\end{tabular*}
\end{table}

\textbf{Overall performance.}
Table~\ref{tab:deep_metrics} summarizes the main results, from which we draw three observations.
(1) \name\ achieves the best Recall@10/100 and NDCG@10/100 on both datasets on all six countries, outperforming all baselines. (2) Token-based GR models (REG4Rec, GenCDR) outperform ID-based models (SASRec, HSTU), showing that tokenization and cross-country parameter sharing provide a stronger backbone, especially for small countries. (3) Single-signal augmentations based only on content or collaboration improve some countries but degrade others, while \name\ combines dual-constrained code-mixing with context-aware reweighting and delivers more consistent gains across all countries.
% Table~\ref{tab:deep_metrics} summarizes the main results, from which we draw three observations. \textbf{(i) \name\ consistently leads on both datasets:} it achieves the best Recall and NDCG at both cutoffs, with Avg6 gains of $+3.01\%/+5.27\%$ on Industrial and $+4.02\%/+6.04\%$ on Amazon over the strongest baseline (R@10/100). \textbf{(ii) Token-based GR is a stronger backbone, and cross-country parameter sharing further helps small markets:} Reg4Rec and GenCDR clearly outperform ID-based models (SASRec, HSTU), and GenCDR additionally improves over Reg4Rec on small countries. \textbf{(iii) Single-signal augmentations inevitably hurt some markets, while \name\ resolves the trade-off:} both content-only and collaboration-only substitutions degrade a non-trivial subset of countries, confirming that no single signal transfers stably. By combining dual-constrained code-mixing with context-aware adaptive reweighting, \name\ delivers consistent per-country gains across all six markets.

\noindent\textbf{Ablation study.}
We conduct an ablation study on three components of \name\ in Table~\ref{tab:ablation}: the token-level distance constraint, the side-information similarity constraint, and the context-aware reweighting. The token-level distance constraint contributes the largest improvement, indicating that codebook proximity is crucial for transferring cross-country behavioral structure. The side-information similarity constraint further refines substitutions by enforcing multi-modal coherence. Removing adaptive reweighting leads to clear degradation, showing that it effectively filters samples whose mixed tokens are incompatible with the original context.

% We ablate \name's three components (Table~\ref{tab:ablation}): the \emph{token-level distance constraint} and the \emph{side-info similarity constraint} (jointly defining a valid code-mixing), and the orthogonal \emph{context-aware adaptive reweighting}. Removing the token-level distance constraint causes the largest drop ($-4.59\%$ R@100), confirming that codebook proximity is the primary channel for transferring cross-country behavioral structure; the side-info similarity constraint adds a complementary refinement by enforcing multi-modal coherence on top. Removing adaptive reweighting incurs a $-2.27\%$ R@100 drop, revealing that it effectively filters out samples whose mixed token is semantically incompatible with the original context.

\begin{figure}[t]
\centering
\includegraphics[width=0.8\columnwidth]{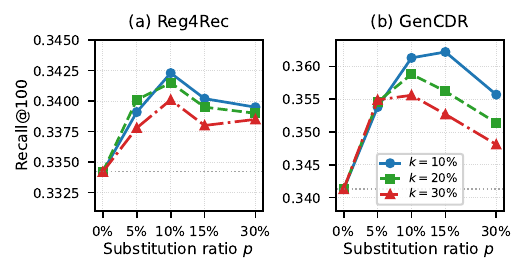}
\caption{Sensitivity to $p$ and $k$ on Industrial (Recall@100, Avg6). Dashed line~=~baseline.}
\Description{Line charts of Recall at 100, averaged over the six countries of the industrial dataset, as the substitution ratio p and the neighbour pool size k are swept on two generative recommendation backbones. Every combination of p and k lies above the dashed horizontal line that marks the un-augmented baseline, and the default setting of p equal to ten percent and k equal to ten percent is close to the best value on both backbones.}
\label{fig:sensitivity}
\end{figure}

\noindent\textbf{Hyperparameter sensitivity and robustness.}
To validate sensitivity and robustness of our method, we sweep $(p,k)$ for our dual-constrained code-mixing on two backbones (Figure~\ref{fig:sensitivity}). On both backbones, all $(p,k)$ combinations consistently outperform the un-augmented baseline, and the default $(p{=}10\%,\,k{=}10\%)$ achieves near-optimal Recall@100 in both cases, confirming that the augmentation is robust to both hyperparameter and backbone choice.

\noindent\textbf{Online Experiments.} We conducted an online A/B test on the industrial platform (10\% live traffic) from May 10 to 20, 2026, covering all six countries. \name\ achieved a consistent increase of \textbf{1.77\%} in \textbf{advertising revenue} and \textbf{2.64\%} in \textbf{orders} on the six-country aggregate, demonstrating significant gains in real-world deployment.
% A 5-day online A/B test on the industrial platform (10\% live traffic, all six countries) yields \textbf{$+1.77\%$ REV} and \textbf{$+2.64\%$ orders} on the 6-country aggregate.
% , with the lift concentrating on smaller markets (B$^M$/C$^S$).

\section{Conclusion}
We presented \name, a cross-country generative recommendation (GR) framework that injects cross-country supervision at the data level through dual-constrained code-mixing. \name\ learns a shared semantic codebook from multi-modal content and behavioral co-occurrence across countries. It then performs token-level code-mixing under content-based and market-aware constraints to synthesize reliable mixed-country sequences without requiring item or user overlap. A context-aware loss reweights mixed samples by their contextual plausibility. This suppresses noisy substitutions while preserving useful cross-country signals. Extensive experiments on offline datasets and an online multi-country e-commerce platform show that \name\ achieves superior performance, effectively transfers knowledge from large to small countries, and offers practical insights for cross-country generative recommendation.